# Unveiling the Superconducting Ground State of Heusler alloy $Pd_2ZrIn$ via muon spin relaxation and rotation measurement

Kavita Yadav[1,2,3], Anoop M Divakaran[2], Tsunehiro Takeuchi[2], K. Mukherjee[3]

[1] Department of Physics, Hindu College, University of Delhi, New Delhi- 110007, India

[2]Toyota Technological Institute, Hisakata 2-12-1, Tempaku, Nagoya 468-8511, Japan.

[3] School of Physical Sciences, Indian Institute of Technology Mandi, Mandi, Himachal Pradesh 175075, India

**Abstract**

Full Heusler alloys $XInPd_2$ (X= Zr, Hf and Ti) have recently attracted significant attention owing to their symmetry-driven electronic structure and also due to the interplay between disorder and emergent ground states. Within this family, $Pd_2ZrIn$ serves as a unique platform to study the effect of disorder on superconducting pairing. This alloy crystallizes in a cubic $L2_1$ structure with significant B2-type antisite disorder. Electrical resistivity and magnetic susceptibility studies confirm bulk type-II superconductivity with a transition temperature $T_C$ ~ 2.2 K. Zero-field μSR results reveal no evidence of spontaneous internal magnetic fields below $T_C$, confirming the preservation of time-reversal symmetry. Transverse-field μSR spectra show the formation of a vortex lattice, consistent with type-II superconductivity, and the superfluid density is well described by a fully gapped, nodeless *s*-wave state with superconducting gap $\Delta$ (0) ~ 0.34 ± 0.01 meV. Furthermore, the estimated ratio of transition temperature and Fermi temperature ($T_C/T_F$) indicates that this alloy lies within the conventional superconducting regime on the Uemura plot. These results establish $Pd_2ZrIn$ as a weakly coupled, dirty-limit, type-II superconductor; with a fully gapped, nodeless order parameter and preserved time-reversal symmetry.

## I. INTRODUCTION

The interplay between crystallographic disorder and superconducting pairing remains a fundamental problem in condensed matter physics [1-3], as it directly impacts the nature and stability of the superconducting state. Crystallographic disorder also provides a direct means to examine the validity of the Bardeen–Cooper–Schrieffer (BCS) framework [4], beyond ideal crystalline systems. In the conventional picture, Anderson's theorem predicts that weak non-magnetic disorder does not significantly affect the isotropic s-wave superconductivity [5]. However, presence of stronger disorder in the system can substantially modify the normal-state electronic properties by enhancing quasiparticle scattering and reducing the electronic mean free path [2,6]. In this case, the electronic mean free path becomes shorter than the intrinsic coherence length, driving the system into the dirty-limit regime [3], where the disorder renormalizes the superconducting length scales even if the underlying pairing symmetry remains conventional. Thus, in disordered superconductors an important question arises that whether superconductivity remains consistent with a conventional isotropic s-wave state or exhibits signatures of anisotropic or unconventional pairing under strong scattering conditions.

In this regard, full Heusler superconductors have attracted considerable attention due to their ability to host superconductivity alongside structural disorder and diverse electronic environments [7]. Several alloys such as $Ni_2NbAl$, $YPd_2Sn$, and $ZrNi_2Ga$ exhibit bulk superconductivity consistent with weak-coupling BCS behaviour [7-10]. Despite their structural similarity, these systems display noticeable variations in superconducting properties. Such variations have been attributed to differences in atomic ordering and site occupancy [11]. In particular, antisite disorder is a common feature observed in full Heusler alloys and can introduce substantial non-magnetic impurity scattering in the system [12]. For example, in $Ni_2NbAl$ and $Ni_2NbSn$ [7], antisite disorder between Ni and Nb atoms alters the density of states (DOS) at the Fermi level, which directly affects the $T_C$. In the case of $ZrNi_2Ga$ [10], deviations from ideal site occupancy have been shown to influence both the electron-phonon coupling strength and the superconducting gap characteristics.

Recent studies also indicate that Heusler superconductors are not restricted to purely conventional superconductivity. For example, in the non-centrosymmetric full Heusler alloy $LuPd_2Sn$ [13], evidence of multi-band superconductivity and mixed pairing characteristics has been reported, including deviations from conventional behaviour and indications of unconventional superconducting features. These observations suggest that Heusler superconductors can host more complex pairing states under suitable electronic and structural conditions. However, the extent to which such behaviour is intrinsic or influenced by disorder remains an open question.

Pd-based Heusler alloys $XInPd_2$ (X = Ti, Zr, Hf) series [8, 14], provide an ideal platform to address this issue. These alloys crystallize in the cubic $L2_1$ structure but exhibit B2-type antisite disorder between atomic sites. In these alloys, the relevant disorder comes from antisite mixing between the X and In atoms, which can be inferred from the weak superlattice reflections observed in the X-ray diffraction pattern. This type of disorder is expected to introduce substantial non-magnetic impurity scattering in these alloys. Although such disorder does not by itself imply pair breaking in a conventional s-wave superconductor, however it can place the system in the dirty-limit regime and thereby significantly influencing the measurable superconducting response. Among these alloys, $Pd_2ZrIn$ has been reported to exhibit conventional superconductivity at low temperatures [8, 11]. However, the available literature also points towards the presence of sample inhomogeneity in this alloy. Specifically, one study reported the presence of two superconducting transitions [8], whereas in another one, small deviations in the magnetic response are noted, which was attributed to an unidentified impurity phase involving Pd and In [11]. These observations suggest that the superconducting properties reported for $Pd_2ZrIn$ may be sensitive to sample quality and phase purity, motivating further investigation of its true intrinsic superconducting behavior. Furthermore, its superconducting ground state has not been examined at the microscopic level, and the nature of the superconducting gap and pairing symmetry remain unresolved. It is therefore essential to determine whether superconductivity in this system remains fully conventional in the presence of substantial antisite disorder and strong scattering. Addressing these questions requires experimental probes that are sensitive to microscopic superconducting properties beyond bulk thermodynamic measurements. Muon spin relaxation and rotation (μSR) is a powerful technique in this respect [15,16]. However, μSR does not directly probe structural disorder itself; rather, it determines the nature of the superconducting state which develops in the disordered alloy. Zero-field μSR measurements allow detection of extremely small internal magnetic fields and provides the direct evidence for time-reversal symmetry breaking [15]. Transverse-field μSR measurements probe the vortex-state field distribution and yield the magnetic penetration depth, enabling determination of the temperature dependence of the superfluid density and superconducting gap structure [16-18].

In this work, we present a combined bulk and microscopic investigation of the superconducting state in the full Heusler alloy $Pd_2ZrIn$. Magnetization, electrical transport, and heat capacity measurements establish the presence of bulk superconductivity in a cubic $L2_1$ structure with pronounced antisite disorder. Zero-field μSR measurements show no observable change in the relaxation rate across the superconducting transition, ruling out the presence of spontaneous internal magnetic fields and confirming the preservation of time-reversal symmetry. Transverse-field μSR measurements demonstrate the formation of a well-defined flux-line lattice and allow direct determination of the superfluid density, whose temperature dependence is consistent with a fully developed, nodeless superconducting gap. The combined results provide a coherent microscopic and macroscopic description of the superconducting state and establish $Pd_2ZrIn$ as a conventional dirty-limit disordered superconductor within the Heusler family.

## II. EXPERIMENTAL METHODS

Polycrystalline $Pd_2ZrIn$ is synthesized by arc melting the high-purity (4N) constituent elements in the stoichiometric ratio under an argon atmosphere. A titanium getter is employed during melting to remove the residual oxygen. To ensure homogeneity, the ingot is flipped and re-melted several times. The final ingot exhibited significant hardness and a negligible weight loss of less than 1%. Phase identification and crystal structure determination are performed at room temperature using powder X-ray diffraction (XRD) with a Rigaku SmartLab rotating anode diffractometer employing Cu-Kα radiation ($\lambda \approx 1.5406$ Å). Elemental analysis of alloy is carried out by employing field emission scanning electron microscope (FESEM, JFEI, USA, Nova Nano SEM). Magnetic measurements of temperature and field dependent DC susceptibility are conducted using a Magnetic Property Measurement System (MPMS, Quantum Design, USA). Heat capacity and electrical resistivity measurements as functions of temperature and magnetic field are performed using a Physical Property Measurement System (PPMS, Quantum Design). For resistivity measurements, the annealed sample is cut into a rectangular shape, and four-point contacts are applied on the top surface using silver paste. Muon spin rotation and relaxation (μSR) experiments are carried out in transverse and longitudinal geometries using the ARTEMIS spectrometer in the S1 area of the Materials and Life Science Experimental Facility (MLF), Japan Proton Accelerator Research Complex (J-PARC), Japan. Prior to the ZF-μSR measurements, the residual magnetic field at the sample position was minimized using the zero-field feedback system of the J-PARC μSR spectrometer, which continuously monitors the local magnetic field and compensates it using three orthogonal Helmholtz coils, thereby establishing a true zero-field condition before data acquisition [19].

## III. RESULTS

### A. Structural Characterization

Fig. 1 (a) represents the room temperature XRD pattern of $Pd_2ZrIn$. It reveals that the alloy crystallizes in a single-phase cubic structure, with no detectable impurity peaks or secondary phases. In full Heusler alloys, the presence of (111) and (200) superlattice reflections along with (220) fundamental peak is an indicator of formation of the $L2_1$ ordered crystal structure [7]. In the present case, similar characteristics are noted in XRD spectra. However, it is observed that the intensity of (111) and (200) reflections are weak in comparison to intensity of (220) reflection [Fig. 1(a)]. This indicates towards the presence of antisite disorder between Zr (Y) and In (Z). Consequently, the Rietveld refinement of the XRD data is also performed using FULL PROF software [20]. The obtained structural parameters are given in Table 1, and a schematic of the corresponding cubic unit cell is shown in Fig. 1(b). During refinement, 50% B2-type disorder is considered between Zr (Y) and In (Z) site. This assumption led to excellent agreement between the observed and calculated XRD patterns, which indicates that the $Pd_2ZrIn$ exhibits substantial B2-type disorder superimposed on the $L2_1$ crystal lattice. Similar feature is noted in other Heusler superconductors [11]. The presence of anti-site disorder can significantly influence the superconducting properties of Heusler alloys. It can alter the electronic band structure of these alloys by broadening the DOS near the Fermi level. In Heusler superconductors, such as $Ni_2NbAl$ or Nb-based alloys, disorder-induced broadening of the electronic states can modify the density of states near the Fermi level and, in turn, influence the electron-phonon coupling strength and the superconducting transition temperature [7, 10]. The presence of antisite disorder is expected to introduce strong non-magnetic impurity scattering and can therefore drive the superconducting state into the dirty-limit regime [7]. Additionally, from EDX analysis, the average composition of the alloy is found to be $Zr_{1.00\pm0.01}Pd_{1.89\pm0.11}In_{0.98\pm0.02}$, which indicates a minor deviation from the nominal composition.To determine how superconductivity is evolved under these conditions in $Pd_2ZrIn$, we have carried out the analysis of temperature and field response of electrical resistivity ($\rho$) and magnetization, which is described below."

### B. Temperature dependent Electrical resistivity

Fig. 2 (a) shows the temperature dependent $\rho$ of $Pd_2ZrIn$, measured from 1.8 to 200 K in $H$= 0 Oe. The figure shows a sharp and drastic decrease in $\rho$(T) near critical temperature $T_C$ ~ 2.2 K with a transition width of 0.1 K, which, along with magnetization and specific heat measurements (discussed in subsequent sections), confirms bulk nature of superconductivity in $Pd_2ZrIn$. Notably, in contrast to the previously reported double superconducting transition in $Pd_2ZrIn$, attributed to annealing-induced sample inhomogeneity [8], the present sample shows a single well-defined transition, indicating improved sample homogeneity and phase purity. The criterion used to determine $T_C$ is the same as that adopted in the magnetization analysis discussed in Section C. The value of residual resistivity ratio (RRR = $\rho$ (200K) /$\rho$ (5 K)) ~1.28, is found to be quite low, indicating disordered nature of the alloy. Such lower values of RRR generally reflects the presence of appreciable static disorder and enhanced elastic scattering in the normal state. This have been previously reported in polycrystalline $YPd_2Sn$ (RRR~ 2.5) and $ZrNi_2Ga$ (RRR ~ 2) Heusler alloys [8,10]. To further examine the low-temperature transport behaviour beyond this disorder contribution, the normal-state resistivity was analysed using the power law ($\rho$ =$\rho_0$ +$A$T$^2$)) fit of the normal state $\rho$ at low temperatures, as shown in Fig. 2(a) using solid red line. Here, the T$^2$ term indicates the contribution of electron-electron scattering to the low-temperature transport. In combination with the Sommerfeld coefficient (γ) obtained from specific heat measurements, the coefficient $A$ further allows an estimate of the Kadowaki-Woods ratio $K_w$= $A/\gamma^2$ [21]. which provides a useful measure for placing the low-energy electronic correlations of $Pd_2ZrIn$ in the context of other intermetallic superconductors. The obtained value of $K_w$ is 12.00 ± 0.01 μΩ-cm mol$^2$ K$^2$/J$^2$ ($A$ = 1.159 × 10$^{-3}$ ± 0.0001 μΩ-cm K$^{-2}$ and $\gamma$ = 9.83 ± 0.02 mJ mol$^{-1}$K$^{-2}$). $K_w$ is found to lie between that noted for heavy fermionic systems and transition metals [22-26]. Similar type of behaviour has also been reported in $LuPd_2Sn$, $Ni_{0.18}Re_{0.82}$ and Ti-V superconductors [13, 27, 28].

Additionally, $\rho$ at different applied magnetic fields are also measured (shown in Fig. 2(b)) to determine the upper critical field ($H_{C2}$), which is discussed in the subsequent section (inset of Fig. 2(b)). Furthermore, carrier concentration is determined by the field-dependent (± 70 kOe) resistivity $\rho_{xy}(H)$ at 10 K (in normal state), as shown in the inset of Fig. 2(a). From the linear fit to the measured data, we have obtained $R_H$ ~ 9.18 ×10$^{-10}$ Ω·m/kOe, indicating that holes are the dominant charge carriers. Using the relationship $R_H$ = 1/$n$e, we have obtained the carrier concentration $n$ = 6.8 ×10$^{27}$ m$^{-3}$, which is characteristic of metallic systems and comparable to other Heusler-type superconductors [11, 29]. Such a high carrier concentration is consistent with a metallic ground state, providing a suitable platform for the emergence of superconductivity. The superconducting properties are further explored through DC susceptibility measurements.

### C. Temperature and field dependent DC susceptibility

Temperature dependence of DC susceptibility ($\chi$) of $Pd_2ZrIn$ measured at $H$= 5 Oe under zero-field cooled (ZFC) and field cooled (FC) conditions is shown in Fig. 3 (a). From the Fig., it can be noted that both ZFC and FC curve exhibits strong diamagnetic signal below $T_C$ ~ 2.2 K, implying the onset of superconductivity and presence of Meissner effect [13, 31, 32]. In contrast to earlier reports on $Pd_2ZrIn$, where small deviations in the magnetic response were attributed to unidentified impurity phases involving Pd and In [11], the present sample shows a well-defined superconducting transition without such anomalous features, indicating improved sample homogeneity. The observed value of $T_C$ is in analogy with previously reported value [8]. Additionally, below $T_C$, a strong irreversibility between ZFC and FC curve is seen, which is one of the key characteristics of type-II superconductors. The superconducting (SC) shielding fraction is also calculated from the ZFC curve and is found to be ~ 75 % at 1.8 K, indicating the presence of bulk superconductivity in the alloy. Fig. 3 (b) represents the field dependent magnetization ($M$ ($H$)) obtained at 1.8 K in the field range ± 0.7 kOe. It exhibits a typical butterfly like magnetic hysteresis curve, which is usually noted in type-II superconductors. In present case, it is not possible to estimate the lower critical magnetic field ($H_{C1}(0)$) due to broadening of virgin $M$ ($H$) curve. The value of $H_{C1}$ is roughly calculated as ~ 50. 00 ± 0.02 Oe at 1.8 K and is described by the magnetic field where the initial slope intersects the extrapolation curve of ($M_{up}$ + $M_{down}$)/2 [13].

The upper critical magnetic field ($H_{C2}(0)$) is also estimated using different measurement techniques: magnetization ($M$ ($T$, $H$)), and $\rho$ ($T$, $H$). In case of $M$ ($T$, $H$), the onset of diamagnetic signal is considered as $T_C$ whereas in and $\rho$ ($T$, $H$), the drop in resistivity value is taken as criteria for $T_C$. The temperature dependence of $H_{C2}(T)$ for both measurements is analysed using the Ginzburg–Landau (GL) model, which describes the linear behaviour of $H_{C2}$ near $T_C$. According to GL theory, the upper critical field is given by [30]:

$$H_{C2}(T) = H_{C2}(0)(\frac{1-t^2}{1+t^2})......... (1)$$

where $H_{C2}(0)$ is the upper critical field at absolute zero and $t= T/T_C$ is the reduced temperature. The fit to eqn. yields $H_{C2}(0)$ = 5.63 ± 0.22 kOe, and 5.87 ± 0.01 kOe from the magnetization (as shown in inset of Fig. 3 (a)) and resistivity measurements (as shown in inset of Fig. 2 (b)), respectively. On application of external magnetic field, two independent cooper pair breaking processes are considered: Pauli limiting field effect and orbital pair breaking effect. The latter effect occurs when the kinetic energy of one of the electrons is increased such that the Lorentz force breaks the cooper pair. The corresponding field is termed as orbital limit field which is given by Werthamer-Helfand-Hohenberg (WHH) model [33]:

$$H_{C2}^{orb} = -\alpha T_C \left.\frac{dH_{C2}}{dT}\right|_{T=T_C} \text{......... (2)}$$

The value of $-dH_{C2}(T)/dT$ in the vicinity of $T_C$ is found to be 2.53 ± 0.02 kOe/K and by considering α = 0.693 (dirty limit superconductors with negligible Pauli limiting and SOC effects), orbital limiting upper critical field ($H_{C2}^{orb}$) is determined as = 3.85 ± 0.03 kOe. In the case of BCS superconductors, the Pauli limiting field is given by $H_{C2}^{p}$= C × $T_C$, where $C$ is 1.86 T/K and is determined to be ~ 40.96 ± 0.03 kOe. The estimated values of $H_{C2}^{orb}$ and $H_{C2}^{p}$ are found to be lower than the experimentally extrapolated $H_{C2}(0)$ values obtained from resistivity and magnetization measurements, as obtained above. This deviation reflects the limitations of the WHH model, likely arising from disorder-driven scattering or multiband effects, and is consistent with the conventional nodeless superconducting state established from μSR and thermodynamic measurements (as discussed in subsequent sections). In addition to this, using the G-L theory, the value of coherence length ($\xi_{GL}(0)$) and penetration depth ($\lambda_{GL}(0)$) of this alloy is estimated [32], which are defined according to following relations as

$$H_{C2}(0) = \frac{\phi_0}{2\pi\xi_{GL}^2}; H_{C1}(0) = \frac{\phi_0}{2\pi\lambda_{GL}^2}\left(ln\frac{\lambda_{GL}(0)}{\xi_{GL}(0)} + 0.12\right) \text{....... (3)}$$

where $\phi_0$ is the magnetic flux quantum and is 2.07 × $10^{-15}$ $Tm^2$. Using above eqns., the values of $\xi_{GL}(0)$ and $\lambda_{GL}(0)$ are calculated to be 243 Å and 1342 Å, respectively. The G-L parameter defined as $\kappa_{GL} = \lambda_{GL}(0)/\xi_{GL}(0)$, is estimated to be = 5.52, indicating the $Pd_2ZrIn$ is weak type II superconductor. The thermodynamic critical field $H_C$ (0) is also evaluated using the relation $H_{C1}(0)$ $H_{C2}(0)$ = $H_{C2}$ (ln $\kappa_{GL}$) and is found to be 403 Oe. These superconducting parameters indicate that $Pd_2ZrIn$ is a type-II superconductor, with characteristics consistent with a weakly coupled superconducting state.

**D. Temperature dependent behaviour of heat capacity**

To elucidate the superconducting behaviour of $Pd_2ZrIn$, temperature dependent specific heat capacity measurements have been carried out in $H$ = 0 Oe as shown in Fig. 4. As is evident from inset (a) of Fig., the specific heat shows a jump at $T_C$ ~ 2.1 ± 0.1 K, determined using the iso-entropic method, which confirms bulk superconductivity in this alloy. With increment in applied $H$, the superconducting anomaly is shifted to lower temperature and completely vanishes at $H$ = 5000 Oe as shown in inset (b) Fig. 4. The application of strong magnetic fields suppresses the superconducting phase at low temperatures, thereby stabilizing the normal state over an extended temperature range. This allows determination of the Sommerfeld coefficient ($\gamma$), the phonon coefficient ($\beta$) and anharmonic coefficient ($\delta$) from standard low-temperature specific-heat fitting given by Debye relation [8]:

$$\frac{C(T)}{T} = \gamma + \beta T^2 + \delta T^4 \text{........... (4)}$$

The specific heat data used for the fitting are measured at an applied magnetic field of $H$= 5000 Oe. We have obtained the values of $\gamma$ = 9.83 ± 0.02 mJ/mole-K$^2$ and $\beta$ = 0.72 ± 0.01mJ/mole-K$^4$ and $\delta$ = 0.29 ± 0.02 μJ/mole-K$^6$ from the fit, represented by the red solid line in inset (c) of Fig 4. The value of $\gamma$ is found to lie in the range observed in Heusler alloys [7, 8, 13]. The above determined values can be used to calculate the normal state parameters of $Pd_2ZrIn$. Sommerfeld coefficient $\gamma$ is associated with DOS at Fermi level [D ($E_F$)] by following formula $\gamma = \frac{\pi^2 k_B^2 \mathrm{D}(\mathrm{E_F})}{3}$, where $k_B$ is the Boltzmann constant ~1.38 × $10^{-23}$ J/K and D ($E_F$) is found to be 4.18 (1) states/eV f.u. According to the Debye model, Debye temperature ($\theta_D$) can be obtained using $\theta_D = \frac{(12\pi^4 RN)^{1/3}}{5\beta}$, where $R$ is the molar gas constant (8.314 J/mol-K) and $N$ is the number of atoms per formula unit and is estimated to be $\theta_D$ ~ 221 ± 0.05 K, which is similar to the $\theta_D$ values for other Heusler superconductors. The Debye temperature $\theta_D$ and $T_C$ are further used to estimate the electron–phonon coupling constant $\lambda_{e-ph}$. The coupling constant is related to $T_C$ and $\theta_D$ through the McMillan formula [34]:

$$\lambda_{e-ph} = \frac{1.04 + \mu^* \ln(\frac{\theta_D}{1.45T_C})}{(1 - 0.62\mu^*)\ln(\frac{\theta_D}{1.45T_C}) - 1.04} \text{....... (5)}$$

where $\mu^*$ represents the screened Coulomb pseudopotential. For conventional superconductors with $T_c$ between $10^{-3}$ and 20 K, $\mu^*$ typically lies between 0 and 0.2. Following Ref. 34, we have used the empirical value $\mu^*$= 0.13. Using $\theta_D$ = 221 ± 0.05 K, and $T_C$ = 2.2 K the electron-phonon coupling constant is estimated to be $\lambda_{e-ph}$ ~ 0.56 suggesting that the alloy is a weakly coupled superconductor. The electronic contribution to heat capacity ($C_{el}$) is determined by subtracting the phonon contribution term from the total heat capacity. $C_{el}$ was normalized by dividing by $\gamma T_C$ as and plotted as a function of the reduced temperature $T/T_C$ shown in inset (d) of Fig. 4. The superconducting transition is characterized by a jump in the electronic heat capacity at $T_C$. The magnitude of the normalized specific heat jump, $\Delta C_{el}/\gamma T_C$ is estimated to be approximately 0.19. This value is considerably smaller than the BCS weak-coupling limit of 1.43 expected for an ideal superconductor [35]. The significantly reduced specific heat jump can be attributed to transition broadening due to disorder, which redistributes the entropy change over a wider temperature range and suppresses the observed discontinuity. Here, we emphasize that the heat capacity data alone do not uniquely establish a disorder-induced modification of the superconducting pairing state. Instead, it provides a thermodynamic support for the conclusion that superconductivity in $Pd_2ZrIn$ develops in a strong scattering environment, which is in agreement with the structural and transport analysis.

### E. Muon Spin Relaxation and Rotation Measurement

**Zero-field μSR:** Zero-field μSR measurements are carried out to probe the presence of any spontaneous magnetic field inside the superconducting phase of $Pd_2ZrIn$, arising from the time-reversal symmetry breaking (TRSB). Zero-field relaxation spectra are acquired at 2.9 K and 0.3 K, corresponding to temperatures above and below $T_C$, respectively. Time evolution of the obtained ZF asymmetry spectra is shown in Fig. 5. The asymmetry spectra do not exhibit any oscillatory components, reflecting the absence of long-range magnetic ordering. This observation is consistent with a superconducting ground state. Additionally, no enhancement in the relaxation behaviour is observed in superconducting state, pointing towards the absence of internal magnetic fields below $T_C$. In the absence of magnetic and electronic moments, the asymmetry decay of muon can occur due to presence of static and random nuclear moments. In this case, the spectra can be modelled using the Gaussian Kubo-Toaybe (KT) function [36] given by

$$G_{KT} = \frac{1}{3} + \frac{2}{3}(1 - \Delta^2 t^2)e^{-\frac{\Delta^2 t^2}{2}} \text{.........} (6)$$

where Δ represents the relaxation rate which originates from the local fields generated by static and randomly oriented nuclear moments. To fit the experimental spectra obtained at 0.3 K and 2.9 K, the following function has been used, which is the product of $G_{KT}$ (t) with an exponential function [37]:

$$G(t) = A_i\, G_{KT}(t) e^{-\Lambda t} \text{......} (7)$$

where $A_i$ and Λ are the initial asymmetry and exponential relaxation rate, respectively. Our ZF-μSR data are well fitted using above eqn. and is shown in Fig. 5. The fitting parameters Δ (nuclear relaxation rate) and Λ (electronic relaxation rate) obtained from the fit are listed in Table 2. The ZF spectra (0.3 K) exhibit no statistical difference compared to the normal state spectra (2.9 K) i.e. the fitting curves of both data show significant overlap. It can also be inferred that the contribution from exponential term is small, and it exhibits temperature independent behaviour across the superconducting transition. Furthermore, nuclear relaxation rate Δ shows no significant enhancement below $T_C$, suggesting the absence of a small spontaneous magnetic field in the superconducting phase. The above observations rule out the possibility of any spontaneous magnetization, thereby indicating that time reversal symmetry is preserved in $Pd_2ZrIn$. The strength of a spontaneous field $|B_{int}|$ can be estimated by $\sqrt{2}\delta\Delta/\gamma_\mu$ . The difference in nuclear relaxation rate above and below the superconducting region, $\delta\Delta$ = 0.0019 ± 0.0002 for $Pd_2ZrIn$, and the muon gyromagnetic ratio $\gamma_\mu$ equals 2π × 135.53 MHz/T. The estimated spontaneous internal field is 0.032 ± 0.001 Oe which is considerably lower than the characteristic fields reported in μSR studies of time-reversal symmetry breaking [38, 39].

**Transverse-field μSR:** The TF-μSR measurements were performed in field-cooled mode down to 0.3 K under applied transverse magnetic field of 100 Oe. Figure 6(a)-(c) shows the TF-μSR asymmetry spectra measured at 0.3 K and 1.8 K in the superconducting mixed state and at 3.3 K in the normal state under an applied field of 100 Oe. At 0.3 K, well below $T_C$, the TF-μSR spectrum shows pronounced relaxation of the oscillatory asymmetry, indicating an inhomogeneous internal magnetic-field distribution associated with the superconducting vortex state. At 1.8 K, closer to $T_C$, the relaxation is considerably weaker, whereas at 3.3 K, above $T_C$, the spectrum exhibits a nearly non-relaxing oscillatory behavior. The observed asymmetry contains contributions from both the superconducting sample and the Ag sample holder. The Ag background has a nearly negligible relaxation rate and therefore contributes a non-relaxing oscillatory component due to muon precession in the applied transverse magnetic field. The sample contribution includes the relaxation due to static nuclear dipolar fields, described by the KT function, and the additional Gaussian depolarization arising from the inhomogeneous magnetic-field distribution of the vortex lattice. The combination of these sample and background contributions gives rise to the observed asymmetry profile, which is not centred symmetrically about zero.

The temperature dependence of the relaxation provides further evidence for the superconducting origin of the additional depolarization. As the temperature increases from 0.3 to 1.8 K within the superconducting state, the superfluid density decreases, leading to a reduction in the second moment of the vortex-state magnetic-field distribution. Consequently, the superconducting Gaussian depolarization rate ($\sigma_{sc}$) decreases, resulting in weaker early-time depolarization at 1.8 K compared with 0.3 K. Upon warming above $T_C$, the vortex-lattice contribution disappears, leaving predominantly the non-relaxing applied-field precession from the sample and Ag background. The superconducting gap structure of $Pd_2ZrIn$ has been investigated by temperature variation of $\sigma_{sc}$, estimated from the second-moment method [42]. To obtain this, the time-domain asymmetry spectra are fitted with [40-44],

$$A(t) = G\ (t) + A_1 \exp\left(-\ \frac{\sigma^2 t^2}{2}\right) \cos\left(\gamma_\mu B_1\, t + \phi\right) + A_2 \cos\left(\gamma_\mu B_2\, t + \phi\right) \ldots\ldots \text{(8)}$$

where $G\ (t)$ is given by eqn. 7 and $A_1$ and $A_2$ are the initial asymmetries of the sample and background, respectively. $\sigma$ is the Gaussian relaxation rate, $\phi$ is the initial phase, and $B_1$ and $B_2$ are the local magnetic field detected by the muons in the sample and in the sample holder (background), respectively. Here, we would like to mention that for temperatures above $T_C$, $Pd_2ZrIn$ is in the normal state, where the superconducting contribution to the muon spin depolarization is absent. Accordingly, the TF-μSR asymmetry spectra are fitted using only a Gaussian damped oscillatory function, without including the relaxation function $G\ (t)$. In contrast, for temperatures below $T_C$, the spectra are analysed using Eq. (8), where the oscillatory component is multiplied by $G\ (t)$ to account for the additional relaxation associated with the superconducting state. The temperature-dependent Gaussian depolarization rate extracted from these fits is then used to determine the superconducting contribution, $\sigma_{sc} = (\sigma^2 - \sigma_n^2)^{1/2}$, from which the temperature dependence of the superfluid density is obtained. The temperature dependence of $\sigma_{sc}$ estimated using the above equation is plotted in Fig. 7. The depolarization exhibits a plateau at low temperature after which it shows decreasing behaviour with increment in temperature reaching zero at $T_C$. This nature can be described by the *s*-wave model in the dirty limit as given by [44]

$$\frac{\sigma_{SC}(T)}{\sigma_{SC}(0)} = \frac{\Delta\ (T)}{\Delta\ (0)}\ tanh\left[\frac{\Delta(T)}{2k_B}\right] \ldots\ldots\ldots \text{(9)}$$

where $\Delta(T)/\Delta(0) = \tanh\ \{1.82(1.018(T_C/T\ -1))^{0.51}\}$ is the BCS approximation for the temperature dependence of the energy gap and $\Delta(0)$ is the magnitude of gap at $T = 0$ K. However, in clean limit it is described by [44]

$$\frac{\sigma_{SC}(T)}{\sigma_{SC}(0)} = 1 + 2\ \int_{\Delta\ (T)}^{\infty} \left(\frac{\delta f}{\delta E}\right) \frac{EdE}{\sqrt{E^2 - \Delta^2(T)}} \ldots\ldots\ldots \text{(10)}$$

where $f$ is the Fermi function and $\Delta\ (T) = \Delta_0\ \delta\ (T/T_C)\ g\ (\phi)$. The term $g(\phi)$ is related to angular dependence of the gap function, where $\phi$ is the azimuthal angle. Depending on the symmetry of the superconducting gap, $g(\phi)$ takes different forms: (i) for *s*-wave; $\phi = 1$, (ii) for *d*-wave; $\phi = \cos(2\phi)$ and (iii) for an anisotropic gap; $(1 + a\mathrm{Cos}(4\phi))/(1 + a)$, where $a$ represents the anisotropic parameter [45]. Among the various models considered, the data is best described by the dirty-limit *s*-wave model, yielding the lowest $\chi^2$ value ($\chi^2_{\mathrm{dirty}} = 1.13$). This is consistent with the relatively low RRR obtained from the temperature-dependent resistivity measurements. The dirty-limit *s*-wave model provides an excellent fit to the data, yielding a superconducting gap of $\Delta(0) = 0.34 \pm 0.01$ meV. This gives the normalized superconducting gap as $\Delta(0)/k_BT_C = 1.726$, showing the weakly coupled nature of the alloy. In contrast to this, the anisotropic model yields an anisotropy parameter $a \sim 0.004$, indicating the presence of isotropic gap. Furthermore, the *d*-wave model also gives a much larger $\chi^2$ value ($\chi^2 = 5.4$), thereby excluding the possibility of *d*-wave symmetry. Hence, the above analysis indicates that the superconducting state in this alloy is fully gapped, isotropic, and consistent with *s*-wave symmetry.

**F. Electronic properties and Uemura Plot**

To gain further insight into the superconducting ground state of $Pd_2ZrIn$, several microscopic parameters are estimated using standard relations for dirty-limit superconductors. Predominantly, the comparison between the electronic mean free path $l_e$ and the BCS coherence length $\xi_0$ is useful for assessing the dirty-limit character of the superconducting state, while the estimate of $T_F$ provides a basis for placing $Pd_2ZrIn$ within the Uemura scheme. The Sommerfeld coefficient $\gamma$, obtained from the low-temperature specific heat capacity data, is related to the quasiparticle number density $n$ through

$$\gamma = \frac{(\frac{\pi}{3})^{2/3} k_B^2 m^* V_{f.u} n^{1/3}}{\hbar^2 N_A} \ldots\ldots (11)$$

where $k_B$ is the Boltzmann constant, $m^*$ is the effective quasiparticle mass, $V_{f.u.}$ is the volume of the formula unit, $N_A$ is Avogadro's number, and $\hbar$ is the reduced Planck constant. The carrier density $n$ is related to the Fermi velocity $v_F$ through

$$n = \frac{1}{3\pi^2}\left(\frac{m^* v_F}{\hbar}\right)^3 \ldots\ldots (12)$$

The electronic mean free path $l_e$ is determined from the residual resistivity $\rho_0$ using

$$l_e = \frac{3\pi^2\hbar^3}{e^2\rho_0 m^{*2} v_F^2} \ldots\ldots (13)$$

For dirty-limit superconductors, the BCS coherence length $\xi_0$ is significantly larger than the mean free path ($\xi_0/l_e >> 1$). In this regime, impurity scattering strongly influences the superconducting properties of the system. Using the experimentally determined values of $\gamma$ and $\rho_0$, these equations were solved simultaneously to estimate the microscopic electronic parameters of $Pd_2ZrIn$. The effective mass is found to be $m^* \sim 2.1\ m_e$, indicating a moderate enhancement of the quasiparticle mass compared to the free electron mass. The carrier density is estimated to be $n \sim 3.4 \times 10^{27}$ m$^{-3}$, while the Fermi velocity is obtained as $v_F \sim 2.4 \times 10^5$ m/s. Using the residual resistivity $\rho_0 = 1.61 \times 10^{-5}\ \Omega$ cm, the electronic mean free path is estimated to be $l_e \sim 25$ Å. The BCS coherence length obtained from the above relations is $\xi_0 \sim 820$ Å, which is much larger than the mean free path, confirming that $Pd_2ZrIn$ lies in the dirty-limit regime. This is in agreement with the analysis of the TF-$\mu$SR depolarization data using a dirty-limit $s$-wave model. The Fermi temperature $T_F$ provides additional insight into the electronic energy scale of the system. For an isotropic spherical Fermi surface, it is defined as$T_F = \frac{\hbar^2 k_F^2}{2m^* k_B}$, where the Fermi wave vector is given by $k_F = (3\pi^2 n)^{1/3}$. Using the estimated carrier density, the Fermi temperature is found to be $T_F \sim 5.2 \times 10^3$ K. The ratio $T_C/T_F$ is widely used to classify superconductors according to the Uemura scheme. In unconventional superconductors, this ratio typically lies within the range $0.01 \leq T_C/T_F \leq 0.1$, whereas conventional BCS superconductors fall well below this band. For $Pd_2ZrIn$, the ratio is estimated to be $T_C/T_F \sim 4.2 \times 10^{-4}$, which is significantly smaller than the unconventional range [46]. This indicates that $Pd_2ZrIn$ belongs to the conventional superconducting regime.

## IV. DISCUSSIONS

The present study establishes the intrinsic superconducting ground state of the full-Heusler alloy $Pd_2ZrIn$ through complementary bulk and microscopic measurements. Earlier studies reported multiple superconducting transitions and anomalous magnetic behavior in this alloy, which has been ascribed to sample inhomogeneity. In contrast to these reports, the present sample exhibits a single superconducting transition as confirmed through magnetization and resistivity measurements. This indicates a homogeneous superconducting state and provides a reliable basis for investigating the intrinsic superconducting properties of $Pd_2ZrIn$.

Structural refinement reveals a substantial B2-type antisite disorder between the Zr and In sublattices, indicating that disorder is an intrinsic feature of $Pd_2ZrIn$. This is supported by the low RRR and the estimated $l_e \sim 25$ Å, which is much smaller than the BCS coherence length $\xi_0 \sim 820$ Å. These observations establish that superconductivity develops in the dirty-limit regime. Importantly, the dirty-limit conclusion obtained from transport analysis is independently confirmed by the TF-μSR superfluid-density analysis, providing a consistent microscopic and macroscopic description of the superconducting state.

The heat capacity measurements further support this picture. Although the normalized specific-heat jump $\Delta C_{el}/\gamma T_C \sim 0.19$ is considerably smaller than the BCS value, the transition occurs close to the experimental base temperature and is significantly broadened. Therefore, the reduced anomaly should not be regarded as evidence for unconventional superconductivity but rather as a consequence of transition broadening and the limited experimental temperature range. The heat capacity results are thus consistent with superconductivity emerging in a strong-scattering environment, in analogy with the structural and transport measurements.

Furthermore, the μSR measurements provide direct microscopic evidence for the nature of the superconducting ground state. Zero-field μSR detects no spontaneous internal magnetic fields below $T_C$, confirming the preservation of time-reversal symmetry. Transverse-field μSR reveals the formation of a vortex lattice, and the temperature dependence of the superfluid density is well described by a dirty-limit isotropic *s*-wave model with $\Delta(0) \sim 0.34 \pm 0.01$ meV. The significantly better agreement of the *s*-wave model compared with anisotropic or *d*-wave models demonstrates that the superconducting order parameter remains fully gapped despite substantial antisite disorder.

Overall, the combined experimental results indicate that antisite disorder in $Pd_2ZrIn$ primarily enhances impurity scattering and modifies the electrodynamic response without altering the superconducting pairing symmetry, consistent with Anderson's theorem for non-magnetic disorder. Furthermore, the small $T_C/T_F$ ratio places $Pd_2ZrIn$ within the conventional superconducting regime of the Uemura classification. These findings establish $Pd_2ZrIn$ as a weakly coupled, dirty-limit, type-II superconductor with preserved time-reversal symmetry and a fully gapped isotropic superconducting state, providing a useful model system for understanding superconductivity in structurally disordered full-Heusler alloys.

## V. CONCLUSION

We have investigated the superconducting ground state of $Pd_2ZrIn$ through a combination of bulk and μSR measurements. The alloy crystallizes in a cubic $L2_1$ structure with significant antisite disorder, which places the system in the dirty-limit regime. Despite this, $Pd_2ZrIn$ exhibits robust bulk type-II superconductivity with $T_C \sim 2.2$ K. Zero-field μSR confirms the preservation of time-reversal symmetry, while transverse-field μSR reveals a well-defined vortex state and a fully gapped superconducting order parameter. The superfluid density is well described by a dirty-limit *s*-wave model, and the extracted gap value supports weak-coupling BCS behavior. These results demonstrate that superconductivity in $Pd_2ZrIn$ remains robust against strong disorder and retains a conventional nodeless superconducting state.

## ACKNOWLEDGEMENTS

The authors acknowledge IIT Mandi for experimental facilities. The muon spin rotation and relaxation (μSR) experiments at the MLF of J-PARC, Japan are performed under Proposal No. 2024A0011. The authors would also like to thank Dr. Jumpei G. Nakamura for his fruitful discussions, valuable comments, and constructive suggestions on the manuscript.

**Tables**

Table 1 Structural parameters obtained from Rietveld refinement of XRD pattern of $Pd_2ZrIn$

| **Lattice parameter (Å)** | 6.569 ± 0.000 | | | |
|---|---|---|---|---|
| **Volume (Å$^3$)** | 283.069 | | | |
| **Wyckoff positions** | **X** | **Y** | **Z** | **position** |
| **Zr** | 0.0 | 0.0 | 0.0 | 4a |
| **In** | 0.0 | 0.0 | 0.0 | 4a |
| **Pd** | 0.25 | 0.25 | 0.25 | 8c |
| **Zr** | 0.5 | 0.5 | 0.5 | 4b |
| **In** | 0.5 | 0.5 | 0.5 | 4b |

Table 2 Parameters obtained from fitting of ZF muon spectra of $Pd_2ZrIn$ using eqn. 7

| ***T* (K)** | **Δ (nuclear relaxation rate)** | **Λ (electronic relaxation rate)** |
|---|---|---|
| 0.3 | 0.2197 ± 0.0023 | 0.0075 ± 0.0001 |
| 2.9 | 0.2171 ± 0.0015 | 0.0077 ± 0.0008 |

**Figures**

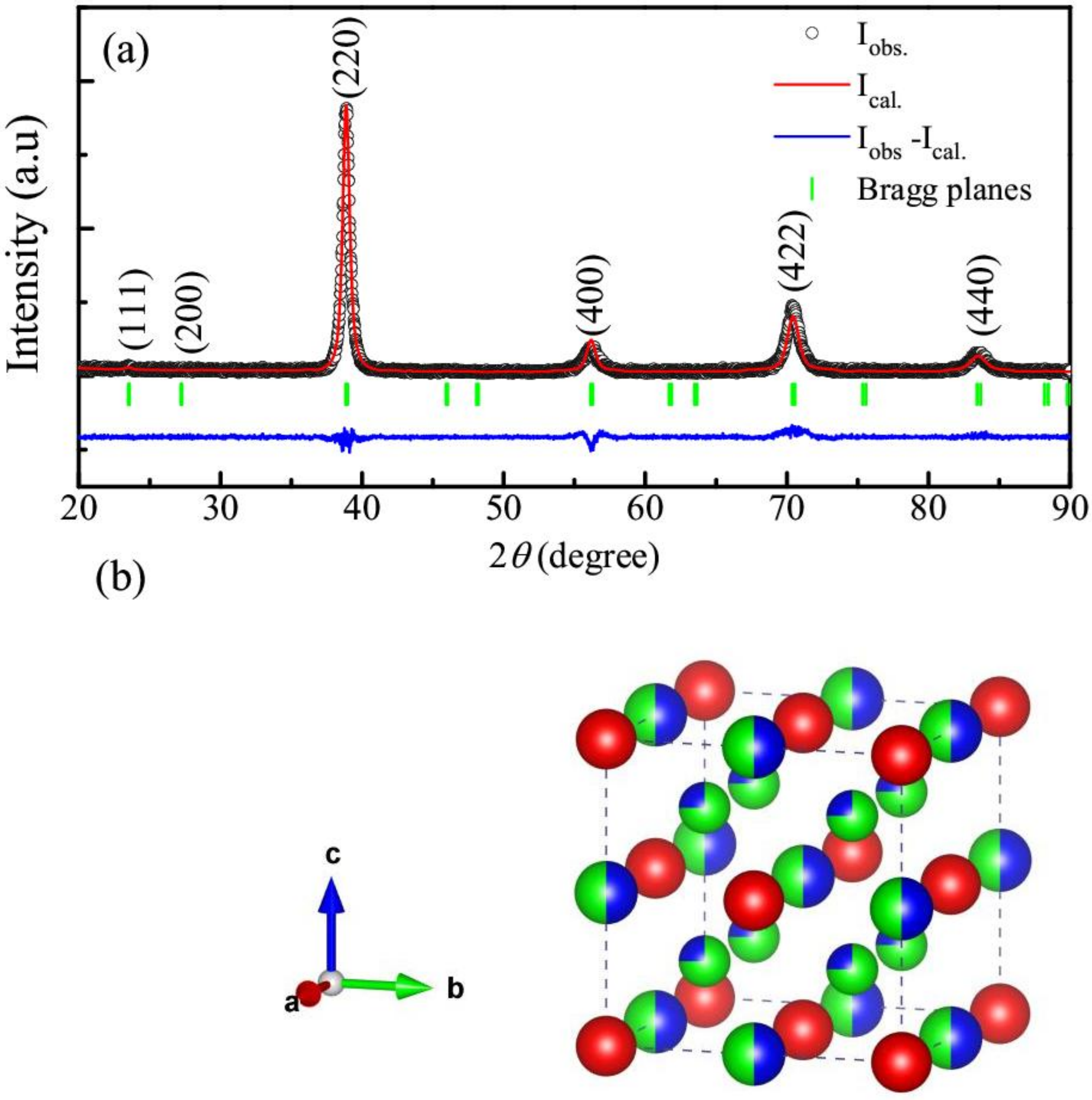

Figure 1 (a) Rietveld refinement analysis of room temperature XRD pattern of $Pd_2ZrIn$ (b) Crystal structure of $Pd_2ZrIn$ (Zr- Red, In- Blue and Pd- Green)

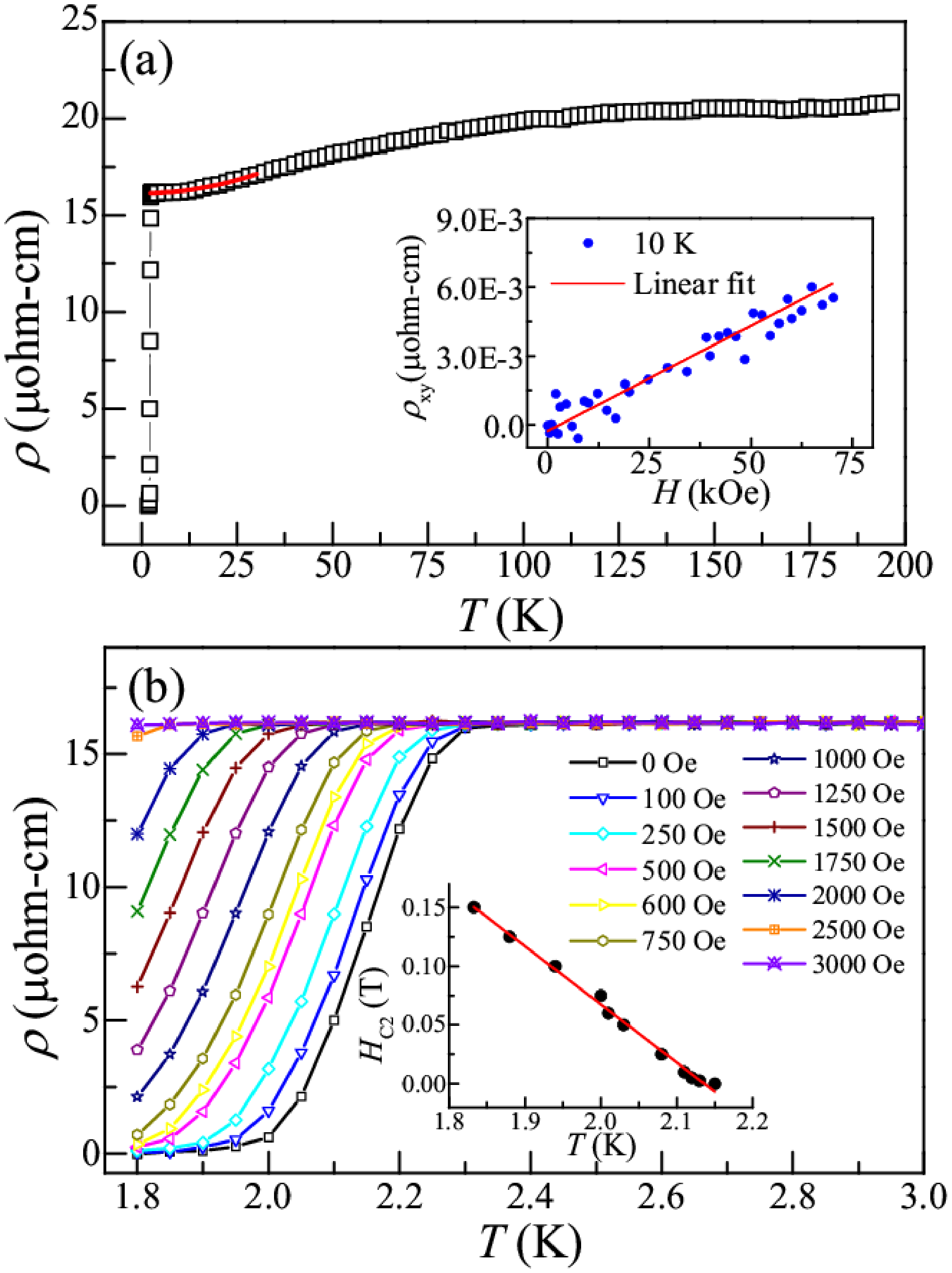

Figure 2 (a) Temperature response of $\rho$ obtained at $H$= 0 Oe in the range 1.8 K to 200 K. Red solid line represents the fit to eqn. $\rho = \rho_0 + AT^2$ Inset: $\rho_{xy}$ obtained at 10 K under applied fields ± 70 kOe. Red solid line represents the linear fit. (b) $\rho$ vs T curves obtained for a range of applied fields. Inset: $H_{C2}$ as function of $T$; Red solid line represents the fit to eqn. 1

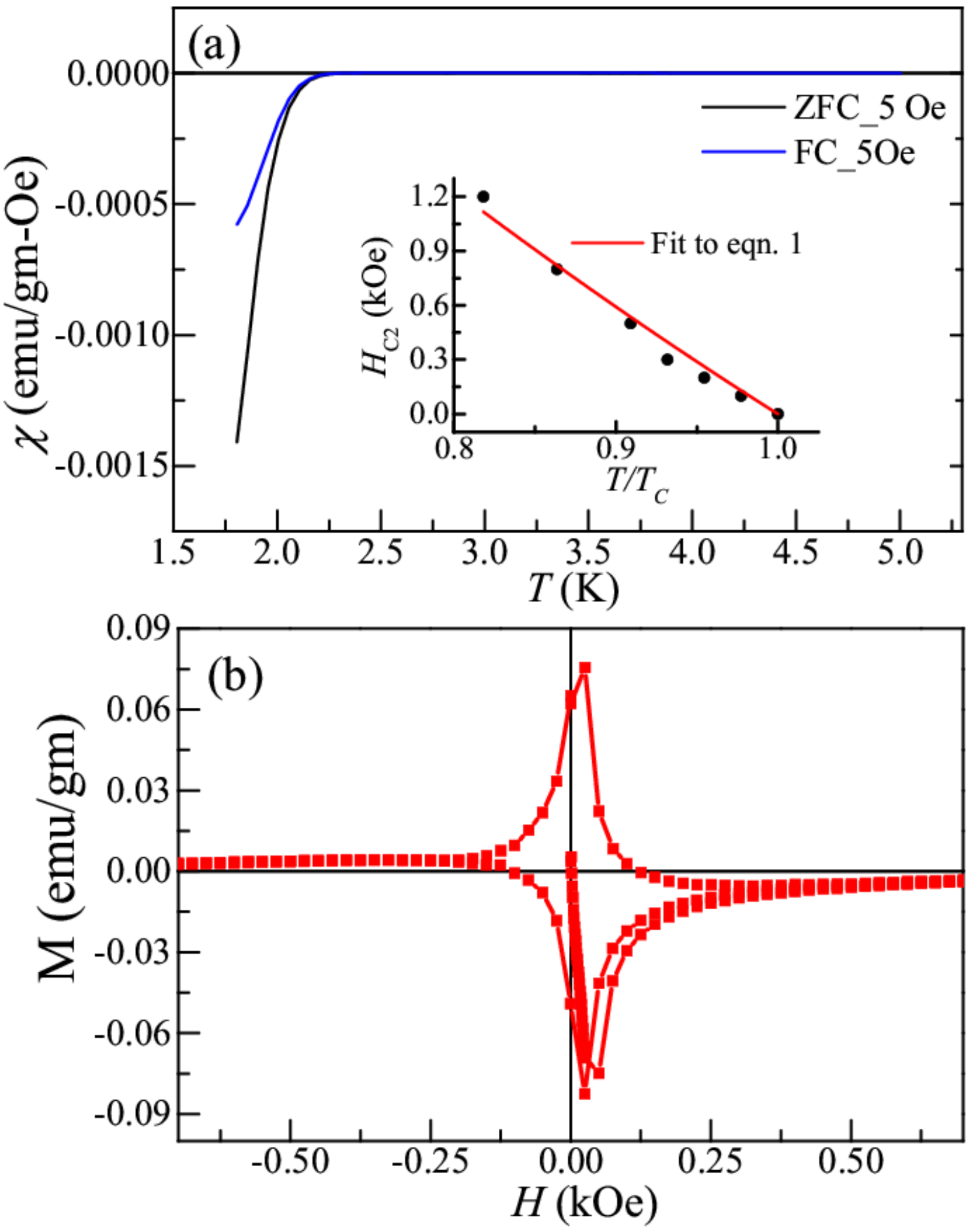

Figure 3 (a) Temperature response of $\chi$ obtained in ZFC and FC condition at $H = 0$ Oe. Inset: $H_{C2}$ as function of $T/T_C$. Red solid line represents fit to eqn. 1. (b) $M$ as function of $H$ obtained at 1.8 K under the applied fields ± 0.75 kOe.

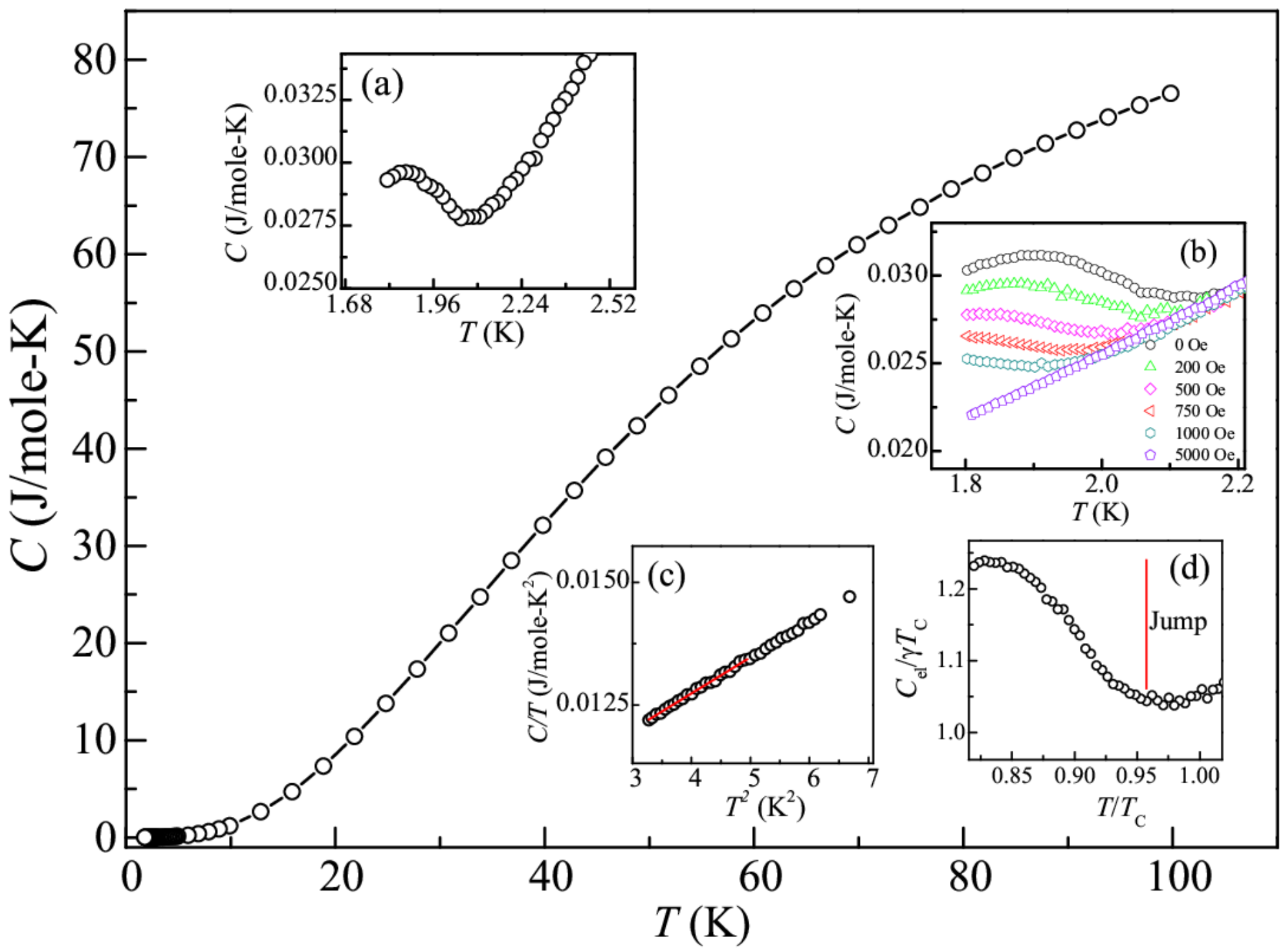

Figure 4 Temperature dependence of $C$ in the temperature range 1.8 K-100 K at $H$= 0 Oe. Inset (a) Same plot in the $T$ range 1.8- 2.8 K. Inset (b) $C$ Vs $T$ at different applied fields Inset (c) $C/T$ vs $T^2$ plot at $H$= 5000 Oe. Red solid line represents fit to eqn. 4. Inset (d) $C_{el}/\gamma T_C$ as a function of $T/T_C$.

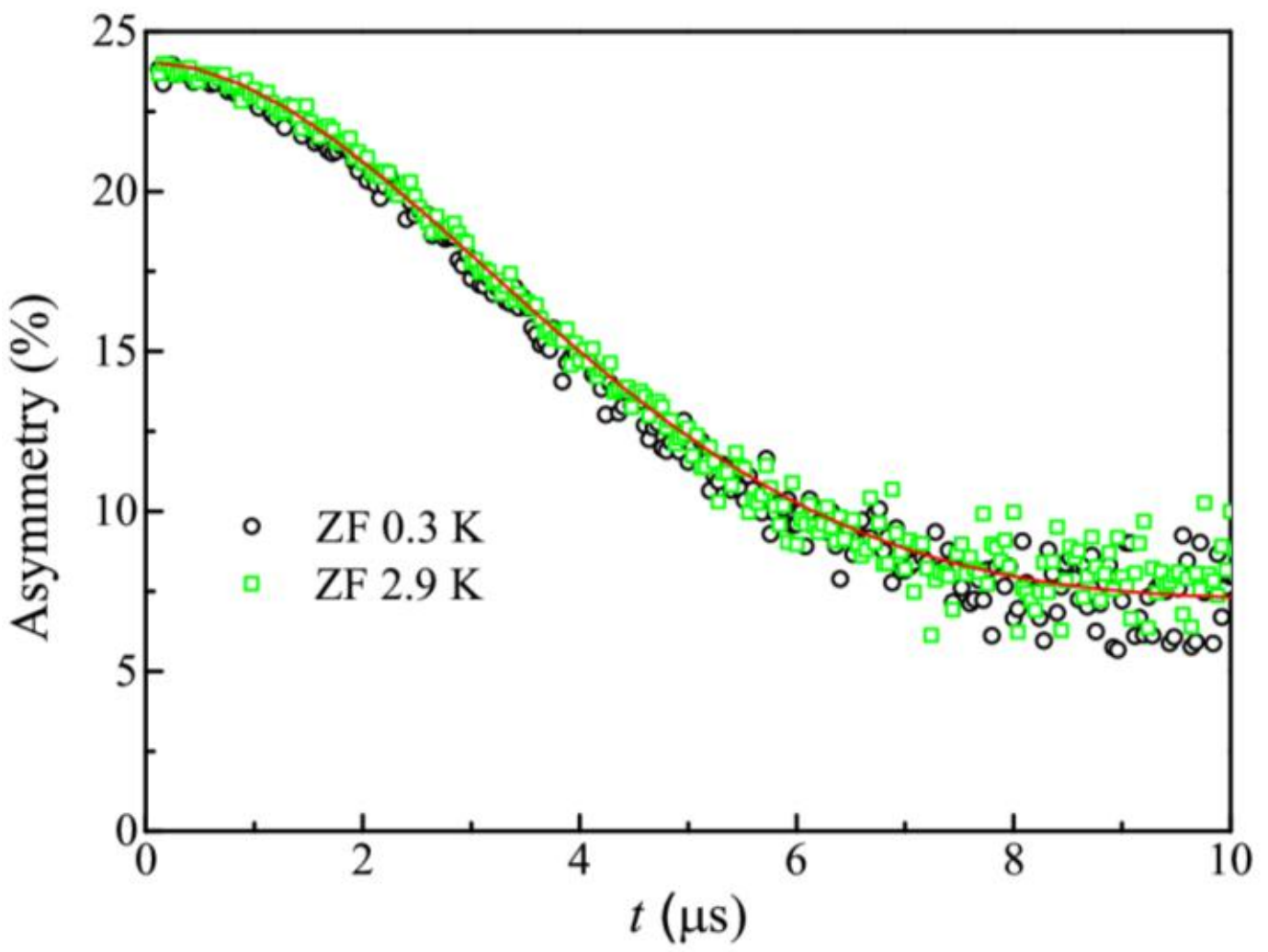

Figure 5 ZF µSR spectra for $Pd_2ZrIn$ at a temperature above $T_C$ (green colour) and below (black colour) $T_C$. Solid red line represents fit to eqn. 7.

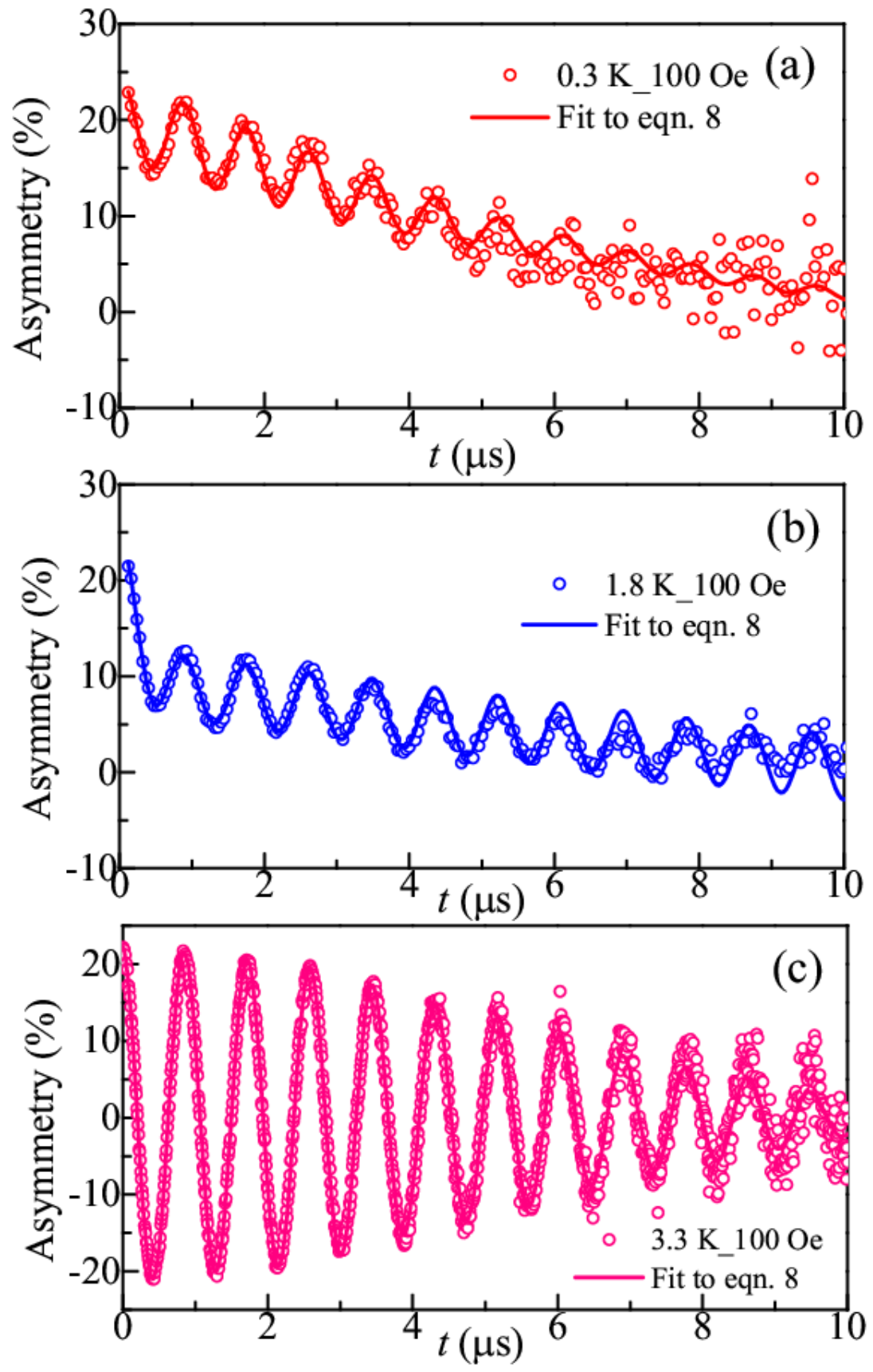

Figure 6 TF- μSR asymmetry spectra (a) 0.3 K, (b) 1.8 K, and (c) 3.3 K at a $H$ =100 Oe, fitted with eqn. 8.

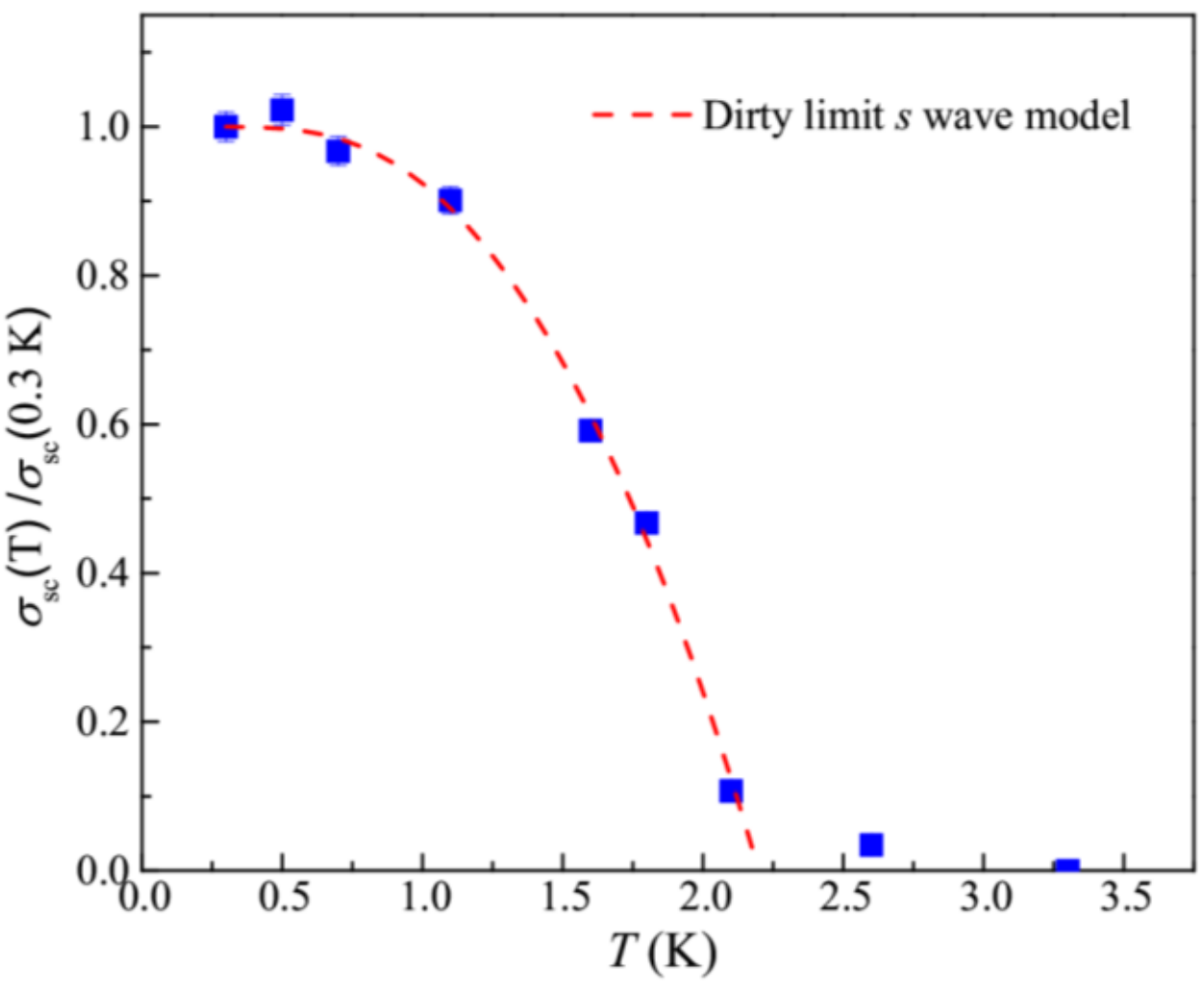

Figure 7 Temperature response of the normalized muon depolarization rate ($\sigma_{sc}$ (T)/$\sigma_{sc}$ (0.3 K)) measured under an applied field of 100 Oe, is well described by a dirty-limit s-wave model.